\documentclass[twocolumn,english,prl,preprintnumbers,amsmath,amssymb]{revtex4}
\usepackage[T1]{fontenc}
\usepackage[latin9]{inputenc}
\usepackage{amstext}

\usepackage{amsmath}
\usepackage{amsfonts}
\usepackage{amssymb}
\usepackage{mathtools}
\usepackage{graphicx}
\usepackage {bm}

\makeatletter
\@ifundefined{textcolor}{}
{%
 \definecolor{BLACK}{gray}{0}
 \definecolor{WHITE}{gray}{1}
 \definecolor{RED}{rgb}{1,0,0}
 \definecolor{GREEN}{rgb}{0,1,0}
 \definecolor{BLUE}{rgb}{0,0,1}
 \definecolor{CYAN}{cmyk}{1,0,0,0}
 \definecolor{MAGENTA}{cmyk}{0,1,0,0}
 \definecolor{YELLOW}{cmyk}{0,0,1,0}
}

\def\be{\begin{equation}}
\def\ee{\end{equation}}
\def\bea{\begin{eqnarray}}
\def\eea{\end{eqnarray}}
\def\bse{\begin{subequations}}
\def\ese{\end{subequations}}

\usepackage{dcolumn}
\usepackage{bm}

\makeatother

\usepackage{babel}
\begin{document}

\preprint{\bibliographystyle{revtex4}}

\title{Fluctuations, renormalizations, and a convective instability in driven wet active matter}

\author{T.R. Kirkpatrick$^{1}$ and J.K. Bhattacherjee$^{1,2}$}

\affiliation{$^{1}$Institute for Physical Science and Technology, University of Maryland, College Park, MD 20742, USA\\
 $^{2}$Department of Theoretical Physics, Indian Association for the Cultivation of Science, Jadavpur, Kolkata 700032, India}

\date{\today}
\begin{abstract}
A scalar model of wet active matter in the presence of an imposed temperature gradient, or chemical potential gradient, is considered. It is shown  that there is a convective instability driven by a (negative) activity parameter. In this non-equilibrium steady state the generic long-ranged correlations are computed and compared and contrasted with the analogous results in a passive fluid. In addition, the non-equilibrium  Casimir pressure or force is computed. Singularities in various physical quantities as the instability is approached are determined. Finally, we give the generalized Lorenz equations characterizing the fluid behavior above the instability and contrast these equations to the Lorenz equations for the Rayleigh-Bernard instability in a passive fluid.

\end{abstract}

\maketitle
In recent years there has been an enormous amount of research on various aspects of active matter \cite{Ramaswamy_2010, Marchetti_et_al_2013}. The hydrodynamic description of active matter rest on identifying the relevant variables, conservation laws, and slow processes, and using symmetry to determine the allowed terms in the equations in some sort of gradient expansion \cite{Brand_et_al_2014}. This is exactly the case for passive matter, but the crucial differences between the active and passive matter description is i) the conservation laws are in general different. For example, active matter, or swimmers, chemically generate their own energy, so that the hydrodynamic energy or temperature equation is not a conservation law \cite{Loi_et_al_2008, Marconi_et_al_2017}, ii) the coefficients in the hydrodynamic equations may be very large compared to their passive counterparts, and even have different signs.

Active matter can be wet, that is, coupled to a momentum conserving solvent, or dry, that is, coupled to momentum absorbing boundaries. Wet active matter is more similar to usual passive fluid hydrodynamics because they have more conservation laws in common. Physically, wet active matter includes bacterial swarms in a fluid, the cytoskeleton of living cells, and biomimetic cell extracts. Many of these objects are approximately spherical objects.

Much of the work on active matter has focused on active liquid crystals. These have either a polar or nematic order parameter that lead to new active terms in the hydrodynamic stress tensor. Depending on the size of the activity, there can be completely new physics such as giant number fluctuations, and spontaneous flow instabilities above an activity threshold \cite{Simha_Ramaswamy_2002, Voituriez_et_al_2005, Narayan_et_al_2007}. Experiments on bacteria swarms and microtubule-based cell extracts \cite{Dombrowski_et_al_2004, Sanchez_et_al_2012} seem to be closely related to numerical simulations \cite{Fielding_et_al_2011, Giomi_et_al_2013, Thampi_et_al_2013} of the active liquid crystal hydrodynamic equations .

Here we will be interested in simpler scalar models of wet active matter, and we will use these models to study a different aspect of wet active matter.  In particular we are interested in active matter in a spatially dependent non-equilibrium steady state (NESS). We compare and contrast the generic long-range correlations that exist in non-equilibrium (NE) passive \cite{Kirkpatrick_Cohen_Dorfman_1982A, Dorfman_Kirkpatrick_Sengers_1994, DeZarate_Sengers_2006} and active matter, and the NE Casimir forces \cite{Kirkpatrick_DeZarate_Sengers_2013, Kirkpatrick_DeZarate_Sengers_2014, Kirkpatrick_DeZarate_Sengers_2016a, Kirkpatrick_DeZarate_Sengers_2016b}. We show that for sufficiently large driving force, or activity, there is a convective instability, which we characterize. To this end, we use a model for wet (momentum conserving) active matter that was introduced by Tiribocchi et.al. \cite{Tiribocchi_et_al_2015}. It is an active fluid version of Model H in the Halperin-Hohenberg classification scheme \cite{Hohenberg_Halperin_1977}. The model has been used to study phase separation in active matter \cite{Tiribocchi_et_al_2015} and to illustrate some general properties of active matter \cite{Nardini_et_al_2017}. 

The hydrodynamic variables in the active fluid version of Model H are a concentration field, $\phi(\bf r, t)$, proportional to the density of active particles, that is \textit {swimmers}, coupled to a momentum conserving solvent. The fluid velocity is $\bf{u}(\bf r, t)$. The equations of motion are,
\begin{equation}
\dot{\phi}+\mathbf{u}\cdot\nabla\phi=D\nabla^2\phi,
\end{equation}
and
\begin{equation}
\dot{\mathbf{u}}+\mathbf{u}\cdot\mathbf{\nabla}\mathbf{u}=-\mathbf{\nabla}p+\nu\nabla^2\mathbf{u}+\mathbf{\nabla}\cdot\mathbf(\Sigma+\mathbf{P})
\end{equation}
Here $D$ is a diffusion coefficient, $p$ is a pressure, which in general is a function of $\phi$ and a temperature $T$, $\nu$ is the kinematic viscosity, and $\mathbf P$ is a Gaussian thermal white noise Langevin force that is specified by it's second moment,
\begin{equation}
\begin{split}
\langle P_{ij}(\mathbf r,t)P_{kl}(\mathbf r',t')\rangle=2k_BT\nu\delta(\mathbf r-\mathbf r')\delta(t-t')\\(\delta_{ik}\delta_{jl}+\delta_{il}\delta_{jk}-\frac{2}{3}\delta_{ij}\delta_{kl})
\end{split}
\end{equation}
There is also a noise term in the concentration equation, but it is not important in what follows. Also in this equation there are higher order gradient terms, as well as non-linearities that we similarly neglect.
$\mathbf\Sigma$ in Eq.(2) is a activity contribution to the stress tensor that is given by,
\begin{equation}
\Sigma_{ij}=-\zeta(\partial_i\phi\partial_j\phi-\frac{\delta_{ij}}{3}(\nabla\phi)^2).
\end{equation}
Such a term is allowed by symmetry in both passive and active fluids. Indeed, historically it is called a non-linear Burnett term \cite{Wong_et_al_1978}. For example, in passive fluids such a term is crucial for understanding the singular behavior of the viscosity as the liquid-gas critical point is approached \cite{Das_Bhattacharjee_2003}. In active matter  there is no obvious restriction on the sign or the magnitude of $\zeta$ \cite{Ramaswamy_2010, Tiribocchi_et_al_2015}. For contractile swimmers ($\zeta<0$) \cite{Williams_et_al_2014, Thung_et_al_2017} the fluid flow increases as the swimmer density gradient increases. This case is of particular interest.

We consider the active fluid in a parallel plate geometry in the $z$-direction with the distance between the plates of size $L$, and the transverse direction $L_{\perp}\gg L$. A spatially dependent non-equilibrium steady state (NESS) is set up by having the plates at a different temperature, or chemical potential. In the former case the imposed temperature gradient will induce a average concentration gradient $\nabla\phi_0$ so that the average pressure gradient is zero. We assume that these average gradients are basically constant, or that there is a linear temperature and concentration profile so  that $\nabla\phi_0=\Delta\phi_0/L$, where $\Delta\phi_0$ is the concentration difference between the two plates. We further assume that we can ignore the dynamical temperature fluctuations since they decay on a relatively fast time scale.

Linearizing the Eqs.(1) and (2) about the NESS allows us to determine the stability of the solution. Assuming no-slip boundary conditions we use the Fourier representations,
\begin{equation}
\nonumber
\begin{split}
 (\delta\phi(\mathbf{r},t), u_z(\mathbf{r}, t))=\frac{2}{L}\sum_{n=1}\int\frac{d\omega}{2\pi}\int_{\mathbf{k}_{\perp}}\\e^{i\mathbf{k}_{\perp}\cdot\mathbf{r}_{\perp}-i\omega t}\sin(\frac{n\pi z}{L})(\delta\phi(\mathbf{k}, \omega), u_z(\mathbf{k}, \omega)). 
 \end{split}
 \end{equation}
Here $\mathbf{k}_{\perp}=(k_x, k_y)$, $\mathbf{k}=(\mathbf{k}_{\perp}, n\pi/L)$,  and $\mathbf{r}_{\perp}=(x,y)$.  The interesting eigenfrequencies are,
\begin{equation}
\omega_{\pm}=\frac{-ik^2}{2}(D+\nu)\pm\frac{i}{2}\sqrt{k^4(D-\nu)^2-4\zeta(\partial_z\phi_0)^2k_{\perp}^2},
\end{equation}
with $k^2=k_{\perp}^2+n^2\pi^2/L^2$. For $\zeta=0$ there is a shear mode and a diffusion mode. For $\zeta>0$ and large, the two modes change from being diffusive to propagating. More interestingly, for $\zeta<0$, and large magnitude, the $\omega_+$ mode becomes unstable. Structurally this is very similar to what happens at the Rayleigh-Bernard (RB) instability. It first occurs at $k_z=k_{\perp}=\pi/L$. The analog of the Rayleigh number is $N\equiv(\Delta\phi_0)^2|\zeta|/D\nu$. Note that physically this is similar to the Rayleigh number, diffusion and viscosity suppress the instability (make $N$ smaller), while the activity parameter takes the place of gravity in driving the instability. The critical $N$ for the instability is $N_c=4\pi^2$. Near the instability,
\begin{equation}
\omega_+\approx -i\frac{2\nu D}{(\nu+D)}([k_{\perp}-\frac{\pi}{L}]^2+\frac{\pi^2}{L^2}\epsilon),
\end{equation}
where $N=N_c(1-\epsilon)$, with $\epsilon\ll1$. Note the implied critical slowing down as the instability is approached.

The feedback mechanism that causes the instability is that a negative (positive) average concentration gradient causes a concentration fluctuation to increase with positive (negative) $u_z$, which in turn causes an increase in the magnitude of $u_z$, etc. Opposing this positive feedback are the viscosity and diffusion. For sufficiently large $|\zeta|(\partial_z\phi_0)^2$ the feedback wins and there is an instability. The linear mathematics of this instability are identical, for example, to the instability in the Richardson combat/arms race model discussed in \cite{Kibble_Berkshire_2004}. If we exclude complex roots, we see below that the non-linear mathematics of the instability are similar to the pitchfork bifurcation that occurs in the RB problem.

It is interesting to compute various equal time correlation functions that characterize the generic long-ranged correlations in the NESS, as well as their amplification as the instability is approached. The largest one for $\zeta<0$ is, in what follows we use units where $k_BT=1$,
\begin{equation}
\langle|\delta\phi(\mathbf{k})|^2\rangle=\frac{(\partial_z\phi_0)^2k_{\perp}^2}{D(\nu+D)k^2}\frac{1}{[k^4-Nk_{\perp}^2/L^2]}
\end{equation}
(for $\zeta>0$ change the sign of $N$). Note that is the absence of activity ($N=0$), this long-range correlation is analogous to the experimentally well verified \cite{Law_et_al_1990} one that appears in a simple fluid
in a temperature gradient \cite{Kirkpatrick_Cohen_Dorfman_1982A}. In the presence of activity it is enhanced (suppressed) compared to the passive fluid result for $\zeta<0$ ($\zeta>0$). Near the instability the singular contribution is,
\begin{equation}
\langle|\delta\phi(\mathbf{k})|^2\rangle_{\mathrm{sing}}\approx\frac{(\Delta\phi_0)^2}{8\pi^2D(\nu+D)}\frac{1}{[(k_{\perp}-\pi/L)^2+\pi^2\epsilon/L^2]}.
\end{equation}

The NE Casimir pressure, $p_{NE}(L)$,  or force is also of interest. For a passive fluid in a NESS it has been discussed in great detail elsewhere \cite{Kirkpatrick_DeZarate_Sengers_2013, Kirkpatrick_DeZarate_Sengers_2014, Kirkpatrick_DeZarate_Sengers_2016a, Kirkpatrick_DeZarate_Sengers_2016b, Aminov_et_al_2015}. Physically, non-linear long-range fluctuations renormalize the pressure \cite{Kardar_Golestanain_1999},
\begin{equation}
p_{NE}(L)=\frac{1}{2}\Big(\frac{\partial^2p}{\partial\phi^2}\Big)_T\overline{\langle\delta\phi(\mathbf{r})^2\rangle},
\end{equation}
where the over-line denotes a spatial average.
For small $|N|$ the equal time correlation function in Eq.(9) is,
\begin{equation}
\overline{\langle\delta\phi(\mathbf{r})^2\rangle}_{|N|\ll 1}\approx\frac{(\Delta\phi_0)^2}{48\pi LD(D+\nu)}
\end{equation}
Near the instability there is a singular contribution given by,
\begin{equation}
\overline{\langle\delta\phi(\mathbf{r})^2\rangle}_{N=N_c(1-\epsilon)}\approx\frac{(\Delta\phi_0)^2}{32\pi^2LD(\nu+D)\sqrt{\epsilon}}.
\end{equation}
For $\zeta>0$ and $|N|\gg 1$ one obtains,
\begin{equation}
\overline{\langle\delta\phi(\mathbf{r})^2\rangle}_{|N|\gg 1}\approx\frac{(\Delta\phi_0)^2}{16LD(\nu+D)\sqrt{|N|}}.
\end{equation}
Again we see that positive activity suppresses fluctuations effects, while negative activity enhances fluctuations.

As the instability is approached, the transport coefficients themselves become singularly renormalized. For example, the mode-coupling renormalization of the diffusion coefficient, $\delta D$ is given by,
\begin{equation}
\delta D=-\frac{\langle u_z(\mathbf{r})\delta\phi(\mathbf{r})\rangle}{(\partial_z\phi_0)}.
\end{equation}
Near the instability the singular contribution is,
\begin{equation}
\delta D\approx\frac{1}{16L(D+\nu)\sqrt{\epsilon}}.
\end{equation}
Numerically this is a very small (it is a $1/L$ effect) perturbation on the bare $D$ unless one is extraordinarily close to the instability. A similar result is obtained for the thermal diffusion coefficient near the RB instability \cite{Kirkpatrick_Cohen_1983}. In practice this means that more sophisticated self-consistent or renormalization group-like treatments are not needed to describe these instabilities \cite{Swift_Hohenberg_1977}.

Finally, we consider the active fluid average NE motion above the instability threshold by constructing a three-mode Lorenz-like model \cite{Lorenz_1993}. The spatial structure of the convection rolls that occur for $N>N_c$ are determined by the critical wavenumbers being $k_{\perp}=k_z=\pi$ (here we use units where $L=1$) and by taking the fluid to be incompressible ($\nabla\cdot\mathbf{u}=0$). Consistent with this we define,
$u_z=A(t)\cos\pi x\sin\pi z$, $u_x=-A(t)\sin\pi x\cos\pi z$, and $\delta\phi=B(t)\cos\pi x\sin\pi z+C(t)\sin 2\pi z$. The equations of motion for a scaled  form of $A(t)$, $B(t)$ and $C(t)$ are,
\begin{equation}
\dot x=\sigma(-x+ry+ryz) \nonumber
\end{equation}
\begin{equation}
\dot y=-xz+x-y 
\end{equation}
\begin{equation}
\dot z=-2z+xy \nonumber
\end{equation}
With $r=N/N_c$ and $\sigma=\nu/D$, the Prandtl number for this system \footnote{This Prandtl number will generally be very large compared to the Prandtl number in the RB problem because particle diffusion coefficients are typically very small compared to thermal diffusion coefficients.}. The non-linear term in the $\dot x$ equation reflects the activity non-linearity and is not present in the Rayleigh-Bernard problem. The fixed points of these Lorenz equations are,
\begin{equation}
x=y=z=0 \quad(\mathrm{stable\quad for\quad} r<1)
\end{equation}
and for $r>1$,
\begin{equation}
\frac{x^2}{2}=r-1\pm\sqrt{(r-1)^2+r-1}\quad (\mathrm{two\quad\ real\quad roots})\nonumber
\end{equation}
\begin{equation}
y=\frac{x}{(1+\frac{x^2}{2})}
\end{equation}

\begin{equation}
z=\frac{xy}{2}\nonumber.
\end{equation}

Compared to the RB Lorenz equations \cite{Kibble_Berkshire_2004, Ott_1993} there are at least three interesting features associated with these Lorenz equations that warrant further study, i) The physical fixed points close to and above the convective instability are $(x,y,z)=(\pm 2^{1/2}(r-1)^{1/4}, \pm 2^{1/2}(r-1)^{1/4}, (r-1)^{1/2})$. In the RB problem, $(r-1)^{1/4}$ is replaced  with $(r-1)^{1/2}$. This implies that the convective transport, which is proportional to $z$, is non-analytic or singular in the control parameter, unlike in the RB problem. ii) The crucial nonlinearity in the $\dot x$ equation is proportional to $r$, and thus increases with driving causing the time independent solution given by Eqs.(17) to become unstable at a smaller $r$ than in the RB problem. This implies that the Lorenz equations for this system are a more realistic representation of the active fluid hydrodynamics than the RB Lorenz equations are for the passive fluid hydrodynamics. iii) There are two additional complex fixed points of the Eqs.(17), compared to the RB problem. This suggest the Hopf-bifurcation and the transition to turbulence in this system will be qualitatively different than in the RB problem.

Indeed, it can be shown that the steady roll fixed point given by the Eqs.(17) will  become unstable to time-dependent flow via a Hopf-bifurcation at a critical $r$, $r_H$, if $\sigma>3$. Typically $\sigma$ will be very large and in this limit a good approximate value for the critical $r$ value is,
\begin{equation}
r_H=1+\frac{\sigma^2+6\sigma+\sigma\sqrt{\sigma^2+16\sigma+24}}{8(\sigma-3)}
\end{equation}
For large $\sigma$ the frequency at the critical point is $\omega\approx\pm\sqrt{8\sigma(r_H-1)}$. We note that at $\sigma=10$, an exact solution of the cubic equation for the Hopf-bifurcation gives $r_H\approx 7.19$ \footnote{Eq. 18 gives $r_H\approx 6.87$ for $\sigma=10$. For $\sigma=50$, both the full cubic equation and Eq.(18) give $r_H\approx16.1$.}. In contrast, for the RB Lorenz equations there is a Hopf-bifurcation if $\sigma>3$ at $r_H=\sigma(\sigma+5)/(\sigma-3)$, so that at $\sigma=10$, this $r_H\approx 21.4$. These two very different values of $r_H$ are consistent with the notion that the Lorenz equations given by Eq.(15) are a better model for the active matter full hydrodynamic equations than the original Lorenz equations are for the RB problem. Physically this is plausible because both sets of Lorenz equations ignore the fluid velocity convective nonlinearity in the $\dot x$ equation, but in the active matter case, this nonlinearity will be sub-leading to the activity nonlinearity if the activity coefficient is large.

We conclude with a number of further remarks:
\begin{enumerate}
\item The presumed size of $\zeta$ here and in \cite{Tiribocchi_et_al_2015} is quite large. In simple passive fluids at liquid state densities we can estimate the scale of $\zeta$ as follows. If $\phi$ is dimensionless then $\zeta$ has the dimensions of $\ell^4/\tau^2=v^2\ell^2$. Here $\ell$ is a length, $\tau$ is a time, and $v$ is a velocity. $\ell$ is the larger of the molecular diameter, $\sigma$,  and the mean-free-path, which for liquid state densities would be $\sigma$, and $v$ is the thermal velocity. For water at STP this would numerically give $|\zeta|\approx 2\cdot 10^{-6}\mathrm{cm^4/sec^2}$, which is about an order of magnitude larger than $D\nu$ in water. The value of $N_c$ and this suggest that the activity part of $|\zeta|$ plays a qualitatively new role when it is larger than the passive one by a factor of $100$ to $1000$.
\item In general in both passive and active fluids a gradient expansion breaks down after Naiver-Stokes order (in simple fluids in three-dimensions) and the generalized description must be non-local \cite{Ernst_Dorfman_1975, Ernst_et_al_1978, Dorfman_Kirkpatrick_Sengers_1994, Belitz_Kirkpatrick_Vojta_2005}. Technically one finds divergences in the calculation of higher order transport coefficients such as $\zeta$. These singular renormalization will have a scale set by the small passive generalized transport coefficients and will presumably not be important. This deserves further study.
\item The transition to turbulence in the Lorenz equations for this system is quite interesting \footnote{J. K. Bhattacherjee and T. R. Kirkpatrick, unpublished}. For $\sigma$ not too large, the Hopf-bifurcation for these equations is backwards. For $r$ close to but below $r_H$ there is a slowly decaying limit cycle about the stable fixed point. For $r>r_H$ there is no stable feature and the fluid motion is turbulent. For larger $\sigma$, the Hopf-bifurcation is forward and there is a stable limit cycle above $r_H$.

In comparison to the RB problem, the smaller values of $(x, y, z)$ at the fixed point, Eq.(17), seem to make the active fluid more stable near $r_H$.
\end{enumerate}

\medskip
Discussions with Jan Sengers are gratefully acknowledged. JKB would like to thank the IPST at the University of Maryland for support during the initial stages of this work. In addition, this work was supported by the National Science Foundation under Grant
No. DMR-1401449.

\end{document}